\documentclass[preprint]{aastex6}
\RequirePackage{epstopdf}
\bibpunct{(}{)}{,}{a}{}{,}

\begin{document}

\received{}
\revised{}
\accepted{}
\slugcomment{Submitted to ApJ}

\shortauthors{Adams et al. }
\shorttitle{Dust Production Rates in the Fomalhaut Debris Disk}

\title{Dust Production Rates in the Fomalhaut Debris Disk from SOFIA/FORCAST Mid-infrared Imaging}

\author{J. D. Adams\altaffilmark{1}, T. L. Herter\altaffilmark{2}, R. M. Lau\altaffilmark{3}, C. Trinh\altaffilmark{1}, M. Hankins\altaffilmark{2}}

\altaffiltext{1}{Stratospheric Observatory for Infrared Astronomy, Universities Space Research Association, NASA/Armstrong Flight Research Center, 2825 East Avenue P, Palmdale, CA 93550 USA}
\altaffiltext{2}{Department of Astronomy, Cornell University, Space Sciences Bldg., Ithaca, NY 14853 USA}
\altaffiltext{3}{Division of Physics, Mathematics and Astronomy, California Institute of Technology, 1200 E. California Blvd., Pasadena, CA 91125 USA}

\begin{abstract}
We present the first spatially resolved mid-infrared (37.1 $\mu$m) image of the Fomalhaut debris disk. We use PSF fitting and subtraction to distinctly measure the flux from the unresolved component and the debris disk. We measure an infrared excess in the point source of $0.9 \pm 0.2$ Jy, consistent with emission from warm dust in an inner disk structure \citep{su16}, and inconsistent with a stellar wind origin. We cannot confirm or rule out the presence of a pileup ring \citep{su16} near the star. In the cold region, the 37 $\mu$m imaging is sensitive to emission from small, blowout grains, which is an excellent probe of the dust production rate from planetesimal collisions. Under the assumptions that the dust grains are icy aggregates and the debris disk
is in steady state, this result is consistent with the dust production rates predicted by \citet{kenyon08} from theoretical models of icy planet formation. We find a dust luminosity of $(7.9 \pm 0.8) \times 10^{-4}$ L$_\odot$ and a dust mass of  8 -- 16 lunar masses,  depending on grain porosity, with $\sim 1$ lunar mass in grains with radius 1 $\mu$m -- 1 mm. If the grains are icy and highly porous, meter-sized objects must be invoked to explain the far-IR, submm, and mm emission. If the grains are composed of astronomical silicates, there is a dearth of blowout grains \citep{pawellek14} and the mass loss rate is well below the predicted dust production values.
\end{abstract}


\section{INTRODUCTION}

Debris disks around stars are thought to be formed from a collisional cascade that occurs when planets form and stir planetesimals. Such collisions continually produce dust around the star. Smaller dust grains are blown out of the system by radiation pressure, while larger grains are pulled into the inner regions of the system by the Poynting-Robertson effect. However, the dust is replenished by the ongoing collisions which occur over the lifetime of the star. \citet{kenyon08} have shown that the range of dust production varies with stellar mass, primordial disk mass, and stellar age. As protoplanets form around 1 -- 3 M$_\odot$ stars, dust production rises rapidly over 1 -- 10 Myr, reaches a peak, then declines over the lifetime of the star. The peak dust production rates for 1 -- 3 M$_\odot$ ranges from $\sim 10^{19}$ to $\sim 10^{22}$ g yr$^{-1}$. More massive disks exhibit peak dust production earlier ($\sim 3$ Myr) than less massive disks ($\sim 10$ Myr). For 1 -- 3 M$_\odot$ stars with ages $> 30$ Myr,  the range of predicted dust production rates narrows and becomes verifiable with observations. Specifically, the detection of small grains, coupled with calculations of their blowout timescales, can be used to determine the dust production rate in a debris disk that is in steady state.

Fomalhaut ($\alpha$ PsA) is a nearby (7.7 pc, van Leeuwen \citeyear{vanleeuwen07}), young (440 Myr, Mamajek \citeyear{mamajek12}) A3V star with an infrared excess and an active debris disk containing a cold submillimeter ring \citep{holland03} that is interpreted as a Kuiper Belt analog. It also has a candidate planetary object, Fomalhaut b \citep{kalas13}. The disk was first discovered as infrared excess in IRAS observations \citep{aumann85}.  The infrared excess is known to originate very close to the star, as seen in Spitzer/IRS spectra \citep{stapelfeldt04} and near-IR interferometric imaging \citep{absil09}.  This excess has been interpreted as having a stellar wind origin \citep{acke12} or as signature of hot dust in the inner disk. \citet{lebreton13} proposed that the infrared excess arises from a dust-producing belt at 1 -- 2 AU from the star, with an inner hot ring of dust located at 0.1 -- 0.3 AU. However, this model overpredicts the mid-IR flux \citep{su16}. Su et al. (\citeyear{su13}) suggested that the dust arises from an asteroid belt, with the near-IR excess originating in hot, magnetically trapped nano grains near the star. The proposed asteroid belt lies near the ice-line at 8 -- 15 AU \citep{su16}, resulting in the production of dust, a drag-in disk from the Poynting-Robertson effect, and a pileup ring due to density enhancement near the silicate sublimation radius.

The cold belt is well-defined \citep{acke12, macgregor17} and exhibits apocenter glow \citep{macgregor17} due to the eccentricity of the system. The disk has been observed in the mid-infrared by \citet{stapelfeldt04}, who used Spitzer/MIPS to obtain images of the disk at 24, 70, and 160 $\mu$m. At 24 $\mu$m, the ansai of the disk were detected; however, the Spitzer observations lacked the spatial resolution to completely resolve the disk from the star. Recently, \citet{acke12} used Herschel/PACS \citep{poglitsch10} and Herschel/SPIRE \citep{griffin10} to image the disk in broad passbands at 70 -- 500 $\mu$m. Although the disk is resolved at 70 $\mu$m, the Herschel data are sensitive to large grains, and are less sensitive to the small grain population. 

An estimate of the dust production rate requires knowledge of the dust grain composition. Fomalhaut shows an asymmetry in scattered light \citep{kalas05}, which could be interpreted as backscattering from large grains \citep{min10}.
Fluffy aggregates have been considered to explain the observed scattering and thermal emission components to the dust SED \citep{acke12}. These grains are similar to icy, porous cometary debris seen in the solar system \citep{fraundorf82}. Using fluffy aggregates, \citet{acke12} derived a mass loss rate of $2 \times 10^{21}$ g yr$^{-1}$ for 1 -- 13 $\mu$m grains in the cold component. However, this value is uncertain due to sparse data in the mid-IR.

An alternate interpretation of the Herschel data was present by \citet{pawellek14}, who used Herschel data and astronomical silicate \citep{draine03} models to examine the grain size distribution in the spatially resolved component. No grains smaller than the blowout size were found, indicating that, if the grains are silicates, the smallest grain size formed in collisions is larger than the blowout size or the grains are rapidly blown out. These grains are iceless despite their location outside the ice sublimation zone.

Motivated to detect small grains and determine a precise measurement of the dust production rate, we obtained new mid-infrared observations with SOFIA \citep{young12}. In \S 2 and \S 3, we discuss SOFIA observations and results. In \S 4, we discuss dust models and results based on icy aggregate and astronomical silicate compositions. In \S 5, we derive the mass loss rates as a result of stellar radiation pressure. \S 6 contains a discussion of the results for both the point source and the Kuiper Belt analog. Our conclusions are presented in \S 7.

\section{OBSERVATIONS}\label{sec:observations}

We observed Fomalhaut with the FORCAST instrument \citep{herter12} on SOFIA
at a wavelength of 37.1 $\mu$m (direct imaging mode), under SOFIA Program $04\_0064$. Observations were carried out over $\sim 10$ hours on four flights during Observing Campaign OC4-G (Southern Hemisphere deployment) in July 2016 (Table \ref{tab:obs}). FORCAST has a field-of-view of $3.4^\prime \times 3.2^\prime$. The observations were performed with matched chopping and nodding 
(NMC mode) and dithering over a grid of nine positions spaced at 30$^{\prime\prime}$. The off-source fields were chosen to be orthogonal to the apparent major axis of the disk in order to minimize the chop throw, thereby minimizing the amount of coma in the telescope beam. Typical dwell times in each nod beam were approximately 60 s, including chopping inefficiencies. A small number of observations were identified to have been taken under aircraft door vignetting, which occurs when the telescope is near its elevation limit and causes significant background variations.

The raw data were processed using the pipeline described in \citet{herter13}. This pipeline
performs chop and nod subtraction and applies corrections for droop, detector non-linearity, multiplexer crosstalk, and optical
distortion.  The chop, nod, and dither positions were aligned using the centroid of the star as a reference position. We averaged the aligned images (excluding the vignetted data), producing an 
image with an effective on-source exposure times of 24,302 s. The combined signals were individually weighted to account for the difference in signal between the matched beams and the non-matched beams. Finally, we subtracted residual background structure across the image caused by the presence of a bright central source \citep{lau13} by using clipping and a second order polynomial fit to the background across each row.
The rectified plate scale for the final image was
$0.768^{\prime\prime}$ per pixel. The effective RMS noise level was $\sim 0.8$ mJy/pixel.
 
Calibration factors at each level altitude were derived from the average photometric instrument response to standard stars observed during the observing campaign, with corrections for elevation angle and altitude. The calibration factors are for a flat spectrum source. The calibration uncertainty was $\sim 6\%$ (Table \ref{tab:obs}). We estimate a total uncertainty of $\sim 10\%$ resulting from both calibration and flat fielding errors.

\section{OBSERVATIONAL RESULTS}\label{sec:results}

\subsection{Imagery}

In Figure \ref{fig:image} (left panel), we show the processed 37 $\mu$m image of Fomalhaut. The signal-to-noise ratio per pixel is low (2) and therefore we have to rely on integrated signal for flux measurements. In Figure \ref{fig:image} (right panel), we show the processed 37.1 $\mu$m image convolved with a Gaussian kernel to enhance the signal-to-noise ratio. The Gaussian kernel had a FWHM of $6.2^{\prime\prime} \times 6.2^{\prime\prime}$. The disk ansai are visible, with average signal-to-noise ratios of 48 per pixel in the Southeastern ansa and an average signal-to-noise ratio of 25 in the Northwestern ansa. For the subsequent analysis, we use unconvolved data for PSF subtraction and to derive the disk flux.

%as well as a component that is interior to the cold belt \citep{acke12}. A very faint halo at the level of the sensitivity limit is seen outside %the ansai, extending to $\sim 25^{\prime\prime}$, or 190 AU from the star. 

\subsection{PSF subtraction}

In order to subtract the unresolved contribution from the disk flux, we performed PSF fitting on a standard star and the Fomalhaut point source. We constructed a model for the PSF consisting of a modified Airy function for diffraction, a polynomial component in the radial direction, and an elliptical Gaussian function for telescope jitter. The source of the a polynomial component is unclear,  but it is prominent in the first Airy bright ring. A similar approach was used by \citet{su17}, who fitted the FORCAST PSF at 35 $\mu$m using a hybrid PSF based on a model and empirical, standard object PSFs. The model parameters were first adjusted to fit the radial profile of a standard star. We used observations of $\alpha$ Boo at 37.1 $\mu$m, taken during OC4-G, as a standard star reference for PSF fitting. The RMS jitter values were $0.^{\prime\prime}.84$ and $0.^{\prime\prime}55$ in the cross-elevation and elevation directions, respectively. An Airy function for diffraction was modified by a third order polynomial across the first bright ring. To enable us to subtract the PSF from the Fomalhaut image, we slightly adjusted to fit the radial profile of the Fomalhaut point source to account for variability in the PSF with telescope elevation. For Fomalhaut, the RMS jitter values were $1.^{\prime\prime}.1$ and $0.^{\prime\prime}84$ in the cross-elevation and elevation directions, respectively

Figure \ref{fig:psfsubstd} shows an image of $\alpha$ Boo, a radial profile of $\alpha$ Boo and the model fit to the data. Figure \ref{fig:psfsubstd} also shows the image and residuals after the model has been subtracted from the data. The scatter in residuals is comparable to the background noise, indicating clean subtraction. Figure \ref{fig:psfsub} shows analogous PSF fitting and subtraction for our Fomalhaut data. The PSF-subtracted image clearly shows the belt in the cold region.

\subsection{Flux profile}

We generated a flux profile along the apparent major axis of the disk by summing the flux in the direction perpendicular to the apparent major axis. The width of the summed region was $23.^{\prime\prime}9$ and the position angle of the apparent major axis was $156^\circ$ (E of N). Figure \ref{fig:profile} shows this flux profile along the apparent major axis both with and without the PSF subtracted. The cold belt is spatially unresolved in the radial direction due to its narrow width (13AU, MacGregor et al. \citeyear{macgregor17}). The outer edge of the Southwestern ansa is located approximately $17^{\prime\prime}$ (131 AU) and the outer edge of the Northwestern ansa is located approximately $19.^{\prime\prime}3$ (148 AU) from the star (Figure \ref{fig:profile}). Taking into account blurring by the PSF, these results agree well with the semi-major axis of the outer edge of the belt as measured by ALMA ($149 \pm 2$ AU, MacGregor et al. \citeyear{macgregor17}).   The Southeastern section shows emission from an inner disk. A fainter flux level in the northwestern section precluded detection of the inner disk. A faint halo is seen in the flux profile extending $\sim 10^{\prime\prime}$ beyond the edges of the ansai, but it is only marginally detected by FORCAST. In archival Herschel images\footnote{http://http://archives.esac.esa.int/hsa/whsa/} at 70 $\mu$m, this halo extends to a radius of $\sim 60^{\prime\prime}$ (460 AU). 

\subsection{Integrated fluxes}

We used a tilted annulus to sum the flux over the region containing the Kuiper Belt analog and its halo. The annuli covered the 100 -- 450 AU ($13^{\prime\prime}$ -- $58^{\prime\prime}$) region to account for blurring by the PSF and the extent of the halo as seen in archival 70 $\mu$m Herschel images.  The inclination ($65.6^\circ$), position angle ($156^\circ$), and eccentricity (0.12) values were taken from \citet{acke12} and \citet{macgregor17}.  The average background per pixel was measured in a tilted annulus with radius $58^{\prime\prime} - 60^{\prime\prime}$ and subtracted from each coadded pixel. We measured an integrated flux density of  $2.0 \pm 0.2$ Jy in the cold region.

%pileup flux = 0.85 Jy
%no pileup = 0.82 Jy

To compute the flux of the point source, we integrated the fitted model PSF. We applied a 15\% color correction for the contribution from the photosphere. The photospheric flux density is expected to be $\sim 1.15$ Jy \citep{castelli04}. The color correction for unresolved warm dust is negligible. Our final result is $2.25 \pm 0.2$ Jy for a flat spectrum source, which is color-corrected to $2.1 \pm 0.2$ Jy for the point source. This result yields an infrared excess of $0.9 \pm 0.2$ Jy at 37.1 $\mu$m.

\section{DUST MODELING}\label{sec:modeling}

In order to characterize the dust populations in the debris disk, we performed dust modeling using a radiation field generated by the star and the assumption that the dust in the disk is in thermodynamic equilibrium. The luminosity $L_g$ of a single grain of radius $a$ is given by:

\begin{equation}
L_g = \pi a^2 \frac{R_*^2}{r^2} \int F(\lambda) Q(\lambda,a) d \lambda = 4 \pi^2 a^2 \int B (\lambda, T) Q(\lambda,a) d \lambda
\end{equation}  

\noindent where $R_*$ is the stellar radius, $r$ is the distance from the star to the dust grain, $F(\lambda)$ is the stellar flux density at the surface of the star, $Q(\lambda, a)$ is the ratio of the absorption area to the geometric cross-sectional area, and $T$ is the equilibrium temperature of the dust grain.

\subsection{Radiation field}
For the stellar flux density, we used a model stellar atmosphere with an effective temperature of 8,750 K and solar metallicity from \citet{castelli04}. We used an average distance from the star of 140 AU for the cold region. The intervening material is optically thin based on its transparency in visible light images \citep{kalas05}.

\subsection{Grain composition and size distributions}

For the grain composition, we adopted icy aggregate grains as suggested for Fomalhaut by \citet{acke12}. We considered a range of porosity, from compact grains to highly porous grains such as those seen in AU Mic \citep{graham07}.
The icy grains are composed by 
mass of 22.5\% silicates, 30.1\% organics, and 47.4\% water ice \citep{pollack94, kataoka14}, with a mean bulk density of $1.68~\rm{g~cm}^{-3}$. $Q(\lambda,a)$ was computed analytically from the real and imaginary refractive indices and derivations given in \citet{kataoka14} for volume filling factors $f$ of 1.0, 0.5, 0.1, and 0.05.

We also considered astronomical silicate grains as the source of dust emission \citep{draine03}. They are compact, with a bulk density of 3.3 g cm$^{-3}$ and a blowout radius of $\sim 5~\mu$m for Fomalhaut.

The grain size distribution $N(a)$ was modeled as $dN/da \propto a^{q}$. The size distributions for blowout grains and for grains larger than the blowout size evolve differently, so we used a separate size distributions for the cases $a \le a_{blow}$ and $a \ge a_{blow}$.
The limits on the grain sizes are set by $a_{min}$ and $a_{max}$, the minimum and maximum grain sizes, respectively. For very large icy grains ($2\pi a/\lambda~\gg~1$), the absorption coefficients approach $1-0.1f$ and the scattering coefficients approach $1+0.1f$ in the mid- and far-IR. $q$, $a_{min}$, $a_{max}$ are the only free parameters in the dust model. The number of dust grains $N(a)$ and integrated luminosity of the dust populations are constrained by the observed SED.

\subsection{Modeling Results}

For icy grains in the cold region, we computed equilibrium temperatures that fall in the range 48 -- 69 K. Fig. \ref{fig:temps} shows equilibrium temperature $T(a)$ as a function of grain radius $a$. The grains in the halo will be at a slightly lower temperature due to their larger distance from the star than the cold belt at a distance of $\sim 140$ AU, but their contribution to the integrated flux is small.

Figure \ref{fig:outer} shows the SED for the cold region, including data from Spitzer/MIPS at 24 $\mu$m \citep{stapelfeldt04} and outer disk fluxes at 70 -- 500 $\mu$m from \citet{acke12}. We note that the outer disk fluxes at 70 -- 500 $\mu$m from \citet{acke12} were derived from a dust model that was convolved with the Herschel beams and fit to the Herschel images. Such an approach relies on grain composition and density distribution to separate the flux of the outer disk from that of the inner disk.  Figure \ref{fig:outer} also shows submillimeter data at 350 $\mu$m \citep{marsh05} and 450 and 850 $\mu$m \citep{holland03}, as well as the integrated flux at 1.3 mm from ALMA \citep{macgregor17}.

%The fluffy aggregate dust grain population is composed of blowout grains with are blown out by radiation pressure in timescales of $10^2 - 10^3$ yr. We fit the observed SED using $q \approx 3.2$ and with minimum grain sizes less than the blowout size. The blowout size is defined by $\beta = 0.5$, where $\beta$ is the ratio of the radiation pressure force to the graviational force. 

Icy grain model SEDs were fitted to the observed SED by eye ($\chi^2 \approx 0.02$) for volume filling factors $f$ of 1.0, 0.5, 0.1, and 0.05.
Table \ref{tab:outer} summarizes the best-fit model parameters and corresponding results for the cold belt. High porosity grains require very large maximum grain sizes, up to 1.2 m for $f=0.05$. Figure \ref{fig:outer} shows the SEDs for the model cases corresponding to the four values of volume filling factors. All these cases can reproduce the observed SED. Consequently, the existing data set does not effectively constrain porosity. Integrating the model SEDs yields the dust luminosities, while integrating the size distribution and grain masses yields the dust mass. The dust mass values are listed in Table \ref{tab:outer}. The dust luminosity is tightly constrained at $(7.9 \pm 0.8) \times 10^{-4}$ L$_\odot$, and the total dust mass ranges from $5.7 \times 10^{26}$ g for compact grains to $1.2 \times 10^{27}$ g for highly porous grains, or $\sim 8$ -- 16 lunar masses. Uncertainties in the dust luminosity and dust mass were derived by varying the model parameters and examining the change in SED within flux uncertainties (10\% at most wavelengths).

We also fit the observed SED (Figure \ref{fig:outer_all}) with astrosilicate grains using the Debris Disk Simulator \citep{wolf05} for two dust population corresponding to the size ranges $a \le a_{blow}$ and $a \ge a_{blow}$. The results from this modeling are listed in (Table {\ref{tab:outer}). The dust mass in blowout grains can be considered an upper limit because the SED can be fit with a single power law size distribution with $a_{min}=a_{blow}$.The disk mass resulting from an astrosilicate model is $(6.6 \pm 0.7) \times 10^{25}$ g, or about 0.9 lunar masses. Although the estimated dust mass from the astrosilicate grain model is lower than that for icy aggregate grains, the two compositions have similar masses for grains with 1 $\mu$m $\le a \le $ 1 mm (Table \ref{tab:outer}).

%We find a total dust luminosity of $2.2 \times 10^4$ and a dust mass of $1.6 \times 10^{23}$ g in %the collisional cascade and $2.5 \times 10^{25}$ g in the large grain population. The total model %SED has a $\chi^2$ goodness-of-fit value of 0.04. 

%In Figure \ref{fig:inner}, we show the observed SED at 37 -- 500 $\mu$m and the best-fitting model for dust in the inner region. The long %wavelength data is shown as upper limits due to contamination from the outer region from the relatively large beam sizes. The data can be fit %with a single model that is devoid of small grains and has a shallow ($\alpha=-1.1$) power law index. The grain equilibrium temperatures were %$\sim 57$ K. Table \ref{tab:inner} summarizes the model parameters and results for the inner region. We calculated a dust luminosty of $3.0 %\times 10^{-5}~\rm{L}_\odot$ and a mass of $5.6 \times 10^{24}$ g for the inner region. 

\section{DISK MODELING}

\subsection{Grain orbits}

In order to compute the timescales for mass loss in the disk from radiation pressure on dust grains, we consider dust grains in orbit under the influence of gravity, radiation pressure, and centrifugal forces. Poynting-Robertson drag at the cold belt is negligible. We used the equations of motion in \citet{harwit88}, modified by the term for radiation pressure, to compute the orbits of blowout grains as conic sections from the cold belt (140 AU) to the distant edge of the halo (450 AU). The incident radiation that contributes to radiation pressure on a grain is $L_{RP} = R_*^2/r^2 \int F(\lambda) \pi a^2  (Q(\lambda,a) + \langle\rm{cos}~\theta\rangle Q_{sca}(\lambda,a) d\lambda$ \citep{tazaki15}. The scattering efficiencies $Q_{sca}(\lambda,a)$ were calculated from the analytical formulae for scattering of fluffy aggregates from \citet{kataoka14}. Under the assumption of intial Keplerian motion, the time was integrated numerically to find the blowout radius and blowout time $\Delta t(a)$ (defined as the time for the grain to reach a distance 450 AU from the star) for each value of $f$. For a hypothetical sample of small grains ($a \ll a_{blow}$), our calculations agree with an analytical calculation for the blowout timescale from \citet{lebreton13}, which is valid for cases where the radiation pressure dominates over the gravitational force and Poynting-Robertson drag. We note our calculated value of 15 $\mu$m for the blowout size for 25\% porosity differs slightly from that of \citet{acke12} (13 $\mu$m). This difference may be due to their inclusion of iron sulfide in the grain composition. The presence of iron sulfide will increase the grain mass density and result in a smaller blowout size. For blowout grains, $\Delta t(a_{blow}) \lesssim 1100$ yr, depending on grain size. For our minimum grain size of 0.1 $\mu$m, $ \Delta t \approx 230$ yr for $f=1.0$ and $ \Delta t \approx 700$ yr for $f=0.05$.

\subsection{The Dust Production Rate}

The disk mass loss rate due to radiative blowout of small grains can be given as: 
\begin{equation}
\dot{M} = \sum_{a=a_{min}}^{a=a_{blowout}} N(a)m(a) / \Delta t(a)
\end{equation} 
Although larger grains are produced in the collisional cascade, they are removed more slowly than the smaller grains, and consequently they contribute far less to the mass loss rate. Table \ref{tab:outer} gives the value of $\dot{M}$ for each value of $f$. For fluffy aggregates, $\dot{M}$ can be expressed as $(1.2 \pm 0.5) \times 10^{21}$ g yr$^{-1}$, with uncertainties derived from uncertainties in porosity and observed fluxes. 

For a disk containing an ongoing collisional cascade, we can compare the observed mass loss rate to a theoretical dust production rate. In order to make this comparison, we scaled the range of predicted dust production rates for 0.01 -- 1.0 $\mu$m grains from \citet{kenyon08} to our range of grain sizes with $a_{min} \le a \le a_{blow}$ based on the observed grain size distribution. The result yields the predicted dust mass that is produced every year for grain sizes in the range $a_{min} \le a \le a_{blow}$. There is a range of predicted dust production rates to account for variations in initial disk mass. The range of dust production rates corresponding to the observed grain size distribution for each value of $f$ are listed in Table \ref{tab:outer}. For icy grains, the predicted dust production rates agree with the observed mass loss rate. For astrosilicates, the mass loss rate falls below the predicted range of dust production rates.

We can also compare the observed disk mass to the predicted disk mass from a collisional cascade \citep{kenyon08}. Table \ref{tab:outer} lists the mass contained in compact grains with 1 $\mu$m $\le a \le$ 1 mm and the mass in the corresponding range for porous grains. The mass in these grains is 0.83 -- 0.94 lunar masses. This range is a factor of $\sim 2$ less than the predicted mass of $\sim 2$ lunar masses for a 1.7 M$_\odot$ star \citep{kenyon08}.

\section{DISCUSSION}\label{sec:discussion}

\subsection{The Nature of the Central Source}
The nature of the inner region has been discussed by \citet{absil09}, \citet{acke12}, \citet{lebreton13}, and \citet{su13, su16}. \citet{acke12} used near-IR (AKARI and Absil et al. \citeyear{absil09}) and Herschel data to measure the slope of the IR excess from $\sim 1$ to 70 $\mu$m. They measured it as a power law in $F_\nu$ with index -0.8. They proposed this power law index was consistent with a stellar wind origin for the IR excess \citep{panagia75}. If the IR excess originates in a stellar wind, the power law predicts an IR excess of $\sim 0.2$ -- 0.3 Jy at 37 $\mu$m. Near-IR excesses and the Herschel 70 $\mu$m excess flux from the central source source can be interpreted as originating from free-free emission associated with a stellar wind. However, 
ALMA measurements of the central source at 870 $\mu$m \citep{su16} indicate that the submillimeter flux is consistent with emission from a photosphere and not free-free emission. In Figure \ref{fig:ptsrc}, we show the mid-IR SED of the central source with photometry from SOFIA/FORCAST (this work) and the predicted SED for free-free emission \citep{acke12}. The SOFIA/FORCAST photometry further rules out a stellar wind origin for the IR excess.

\citet{lebreton13} proposed that the IR excess in the inner region originates from warm dust near the star at 1 -- 2 AU. Such dust is within the ice sublimation radius. They predict an IR excess of $\sim 0.73$ Jy at 37 $\mu$m, which is slightly lower than the observed value, but still lies within the SOFIA/FORCAST uncertainties. However, these models overpredict the flux at 10 -- 24 $\mu$m \citep{su16}.

\citet{su16} also propose that the IR excess originates from warm dust, but they consider an asteroid belt at the ice line (8 -- 15 AU). Astronomical silicate dust is produced in this belt, with larger grains dragged inwards under Poynting-Robertson drag. The dragged-in grains form a warm disk and might be halted in a pileup ring at 0.23 AU due to density enhancement near the silicate sublimation radius. The presence of a pileup ring can cause additional IR excess by $\sim 0.03$ Jy over the ice-line dust and warm disk. Figure \ref{fig:ptsrc} also shows the SED for warm dust from \citet{su13,su16}. The SOFIA/FORCAST data  support the warm disk hypothesis. However, the SOFIA data do not confirm or rule out the presence of a pileup ring, which results in additional excess of only 30 mJy at 37.1 $\mu$m.

\subsection{Dust Production in the Kuiper Belt Analog}

\citet{acke12} derived a mass loss rate of $2 \times 10^{21}$ g yr$^{-1}$ using fluffy aggregate grains with 25\% porosity under a $q=-3.5$ size distribution. For 25\% porosity, we derive the slightly lower value of $\dot{M} = 1 \times 10^{21}$ g yr$^{-1}$. We attribute the difference in size distributions to additional constraints on the observed SED from Spitzer/MIPS \citep{stapelfeldt04}, SOFIA (this work), and ALMA \citep{macgregor17}.

Highly porous (91 -- 94\%) grains are observed in AU Mic (Graham et al. \citeyear{graham07}) in scattered light. For Fomalhaut, highly porous grains would imply that the size distribution extends to meter-sized objects, if the material is composed of icy aggregates. Similar to mm-sized compact grains, such large objects are optically thick and highly ($> 99\%$) absorbing of radiation. For highly porous grains, the blowout size is $\gtrsim 100~\mu$m and the minimum grain size is $\sim 60~\mu$m. in the Fomalhaut Kuiper Belt analog. Such a size distribution is consistent with the proposed belt of micro-asteroids \citep{min10}.

If the grains are icy aggregates, the difference between the observed dust mass of $\sim 1$ lunar mass and the value of $\sim 2$ lunar masses predicted by \citet{kenyon08} is not unexpected. The predicted dust mass corresponds to a different grain composition and porosity, with different absorption and scattering properties than the icy grains. Consequently, the predicted mass loss rate and accumulated disk mass will differ from the observed values for icy grains.

Icy aggregate dust grains can simultaneously reproduce the observed scattering and long wavelength thermal emission.
If we extrapolate the predicted dust production rate for fine grains to our observed size distribution, we find it to be consistent with the theoretical predictions of icy planet formation \citet{kenyon08}. Therefore, the icy grain hypothesis is consistent with the observed emission and the expected dust production rates. However, we caution that the models of \citet{kenyon08} differ from the Fomalhaut debris disk. The Fomalhaut Kuiper belt analog is narrower ($13.5 \pm 1.8$ AU, MacGregor et al. \citeyear{macgregor17}) than the theoretical disks, which span 30 -- 150 AU, so the theoretical disks have a larger volume of planetisimals that can be stirred by icy planets. In addition, the extrapolation of the dust production rates from fine (0.01 -- 1.0 $\mu$m) grains to larger grains ($a \ge 0.1~\mu$m) relies on the assumption that dust is produced in steady state. An alternate interpretation could be that dust was produced 230 -- 700 yr ago and grains with $a < 0.1~\mu$m have all been blown out of the system. In that case, the predicted dust production rate cannot be extrapolated because the dust production is instantaneous, rather than steady state. The steady state assumption implies that the minimum grain sizes are larger than $\sim 1~\mu$m, or in the case of highly porous grains, larger than $\sim 50~\mu$m. 

Both icy grain models and astrosilicate models can reproduce the observed SED. However, they differ in that the astrosilicate model indicates a dearth of blowout grains. This result confirms the results of \citet{pawellek14}. If the grains are astrosilicates, they are produced at a lower rate or form with a minimum size that is larger than the blowout size.

%For grains with sizes 0.01 -- 1.0 $\mu$m, \citet{kenyon08} predict a range of dust production rates for disks around 1 -- 3 M$_\odot$ stars at an age of 440 Myr. These values are $1.8 \times 10^{20}$ -- $7.9 \times 10^{20}$ g yr$^{-1}$, depending on intial disk mass, which are consistent with icy dust production results (Table \ref{tab:outer}.  

\section{CONCLUSIONS}

We have used SOFIA/FORCAST to observe Fomalhaut at 37.1 $\mu$m. The observations have constrained the nature of the central source and the cold debris disk. The infrared excess of the point source at 37.1 $\mu$m is consistent with the presence of an asteroid belt at the ice-line proposed by Su et al. (\citeyear{su13}, \citeyear{su16}). The 37.1 excess in the point source rules out a stellar wind as the origin of infrared excess in the central source. 

Under the assumptions that the cold Fomalhaut debris disk is composed of icy aggregate dust grains and the disk is in a steady state collisional cascade, our observed mass loss rate is consistent with theoretical predictions for dust production rate from icy planet formation \citep{kenyon08}. Icy grains remain a plausible dust grain type for the Fomalhaut because their presence around Fomalhaut would explain the observed scattered light, thermal emission, and the observed dust production rate.  However, if the dust is composed of astrosilicates, the observed mass loss rate is less than the predicted range of dust production rates.

\acknowledgments
\noindent  We thank K. Stapelfeldt for sharing his PSF-subtracted MIPS image of Fomalhaut and K. Su for sharing her inner disk model SEDs. We thank the SOFIA ground crew, flight crew, and Mission Operations for their successful execution of the SOFIA observations. We also thank an anonymous referee whose suggestions led to the improvement of this manuscript. This work is based on observations made with the NASA/DLR Stratospheric Observatory for Infrared Astronomy (SOFIA). SOFIA science mission operations are conducted jointly by the Universities Space Research Association, Inc. (USRA), under NASA contract NAS2-97001, and the Deutsches SOFIA Institut (DSI) under DLR contract 50 OK 0901. 

{\facility {\it Facilities}: Spitzer, SOFIA, Herschel}

\newpage
\begin{figure}
\plottwo{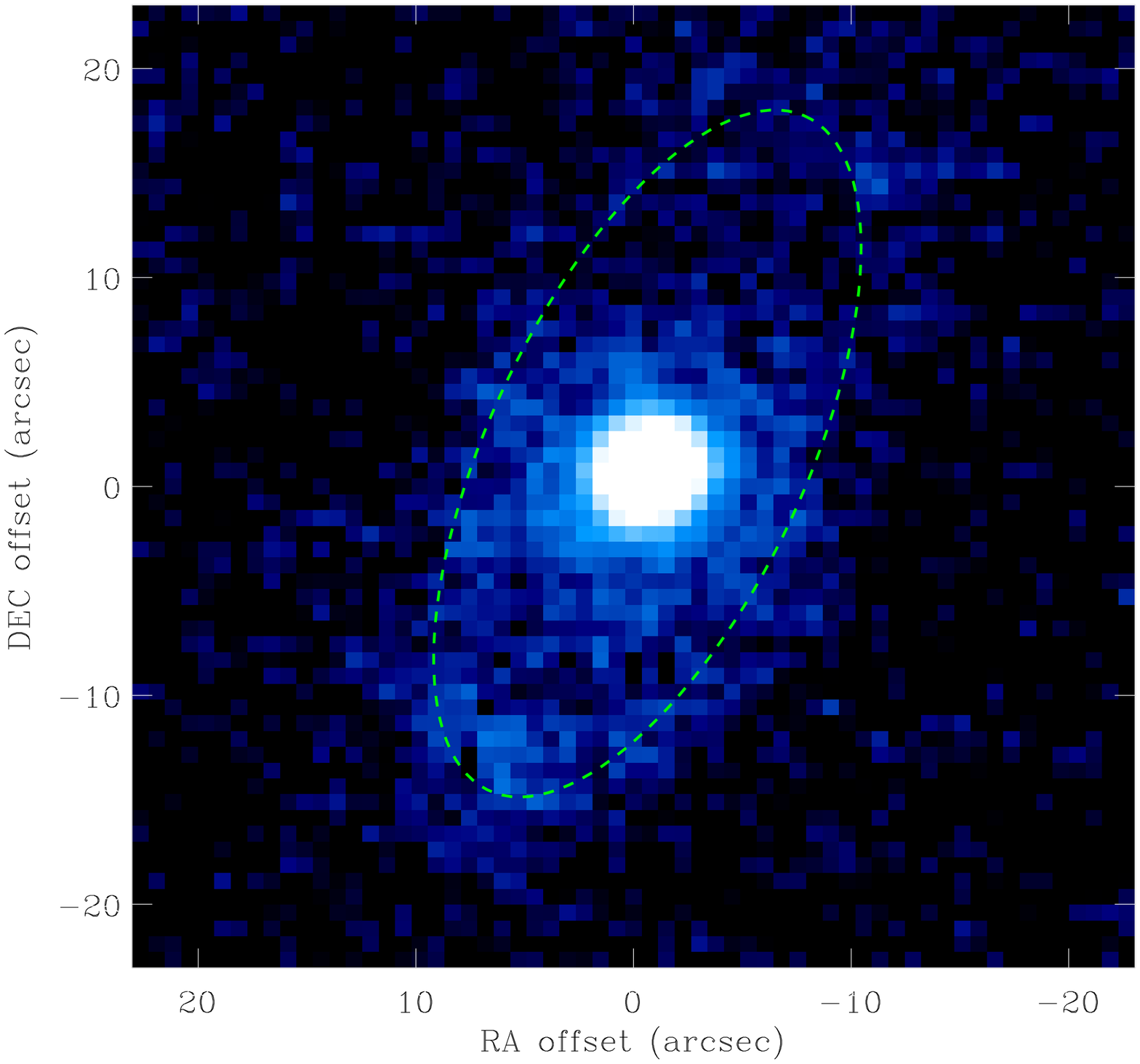}{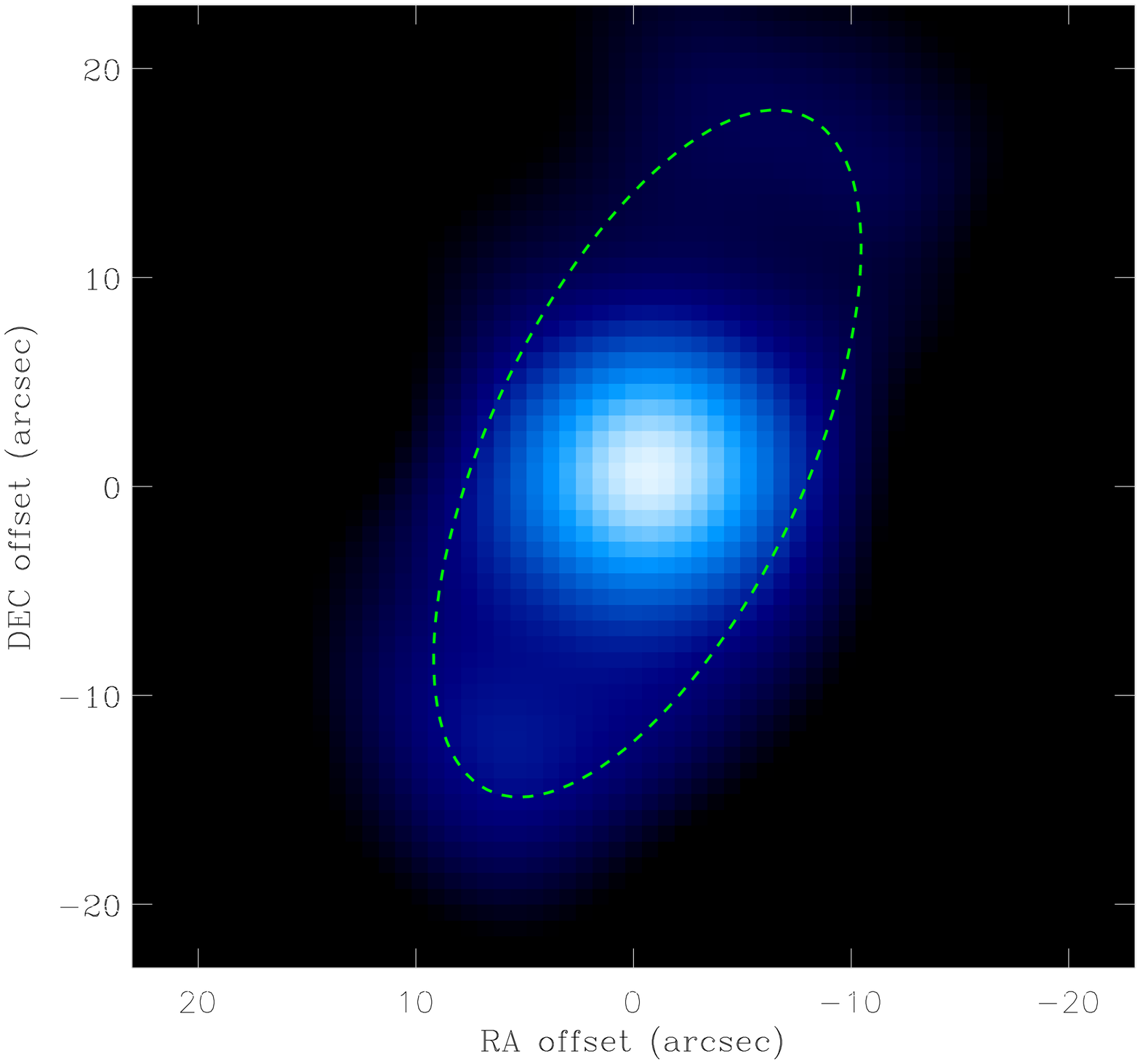}
\caption{{\it Left}: Processed image of Fomalhaut at 37.1 $\mu$m. The dashed green ellipse is the approximate location of the 70 $\mu$m cold belt from \citep{acke12}. {\it Right}: Smoothed image of Fomalhaut at 37.1 $\mu$m with the same cold belt location. The smoothing kernel was a Gaussian with FWHM of $6.2^{\prime\prime} \times 6.2^{\prime\prime}$. \vspace{5.0cm}
\label{fig:image}}
\end{figure}
%\vspace{7.0cm}

\newpage
\begin{figure}
\plotone{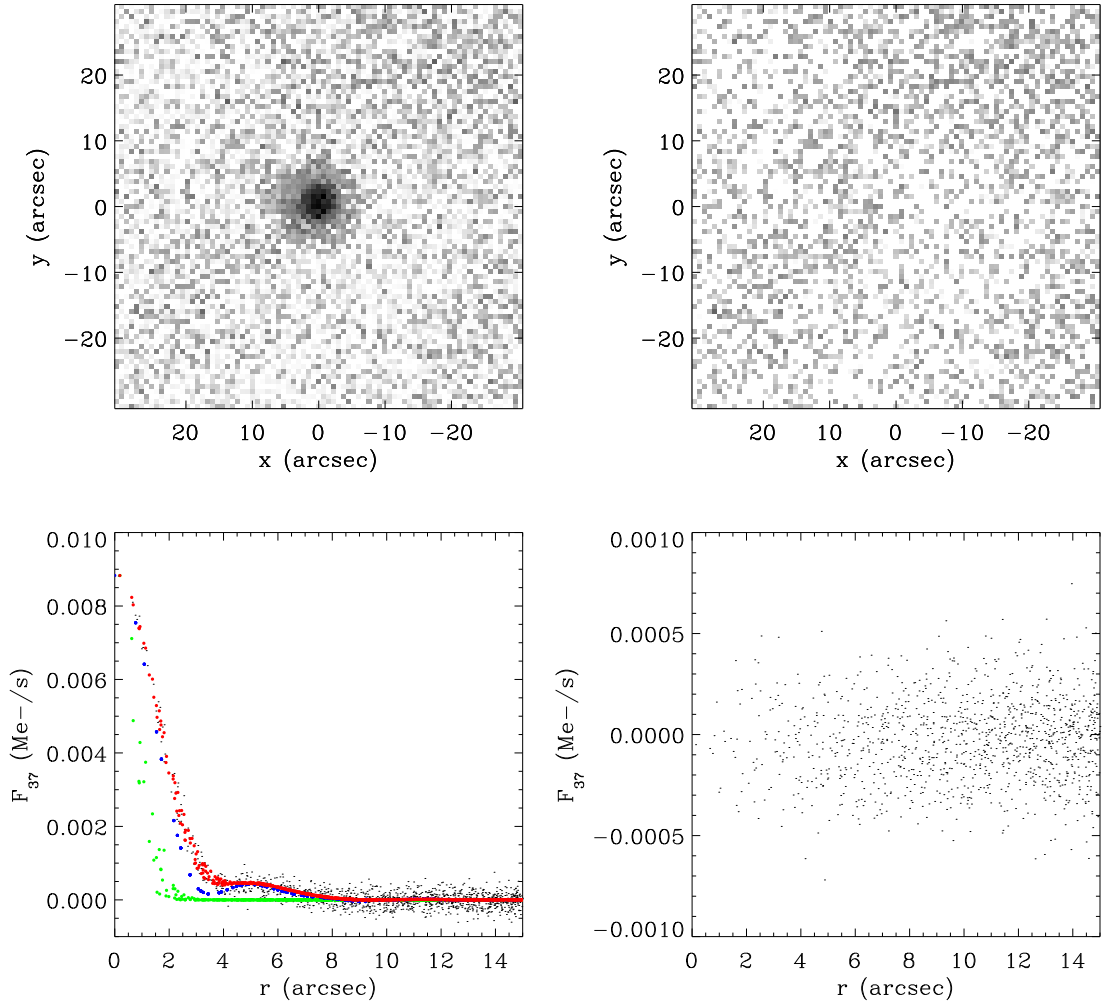}
\caption{PSF fitting and subtraction for $\alpha$ Boo. {\it Upper left}: FORCAST image of $\alpha$ Boo at 37.1 $\mu$m, averaged over the four OC4-G flights. {\it Lower left}: Radial plot for  $\alpha$ Boo (black dots), with flux densities in instrumental units. Model fits to
the data (red points) include contributions from elliptical Gaussian jitter (green points) and a modified Airy function for diffraction and non-Gaussian components (blue points). {\it Upper right}: Image of $\alpha$ Boo at 37.1 $\mu$m after PSF subtraction. {\it Lower right}: Radial plot of residuals after PSF subtraction.
\label{fig:psfsubstd}}
\end{figure} 

\newpage
\begin{figure}
\plotone{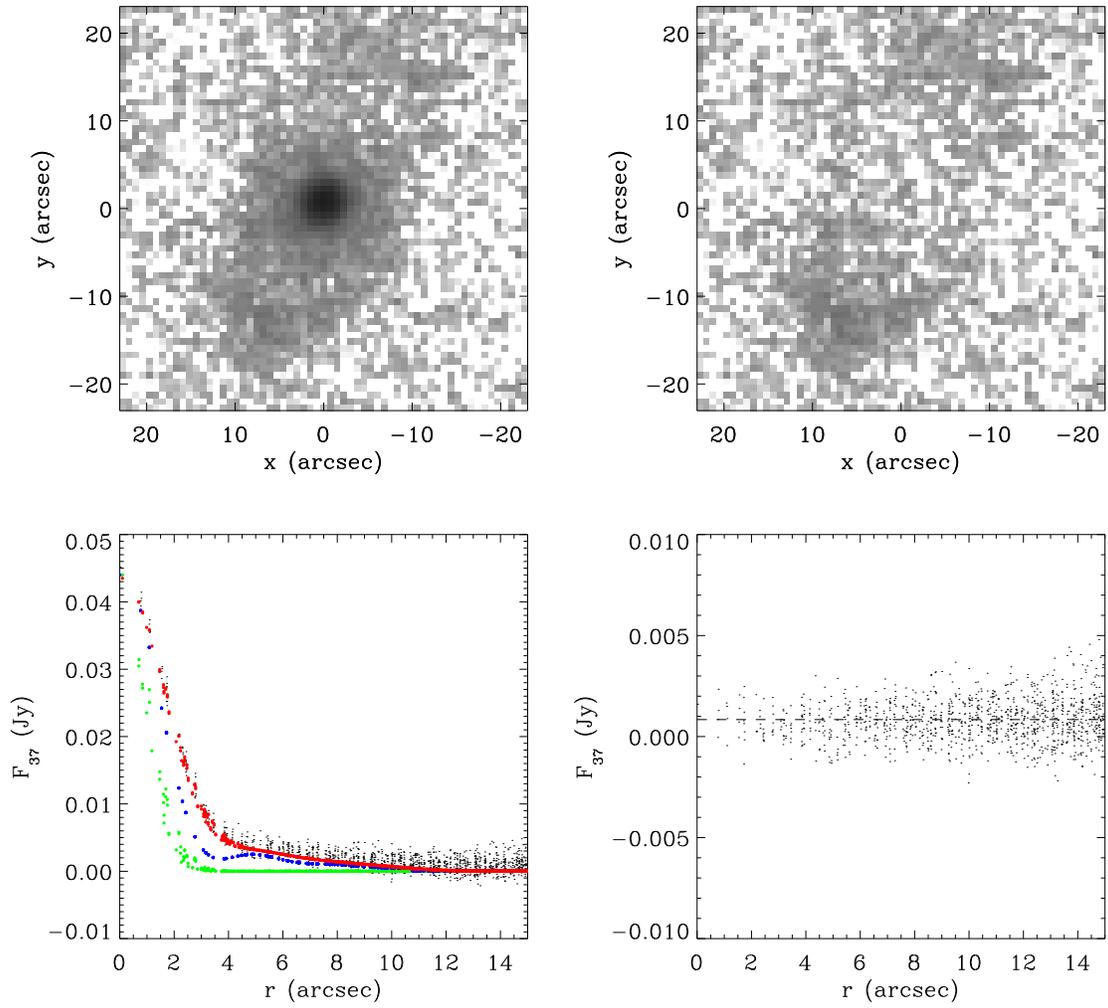}
\caption{Analogous to Figure \ref{fig:psfsubstd}, but for Fomalhaut. The jitter model parameters were adjusted to fit the Fomalhaut data. The Fomalhaut PSF model uses the same modified Airy funtion as the  $\alpha$ Boo PSF model.
\label{fig:psfsub}}
\end{figure} 

\newpage
\begin{figure}
\plotone{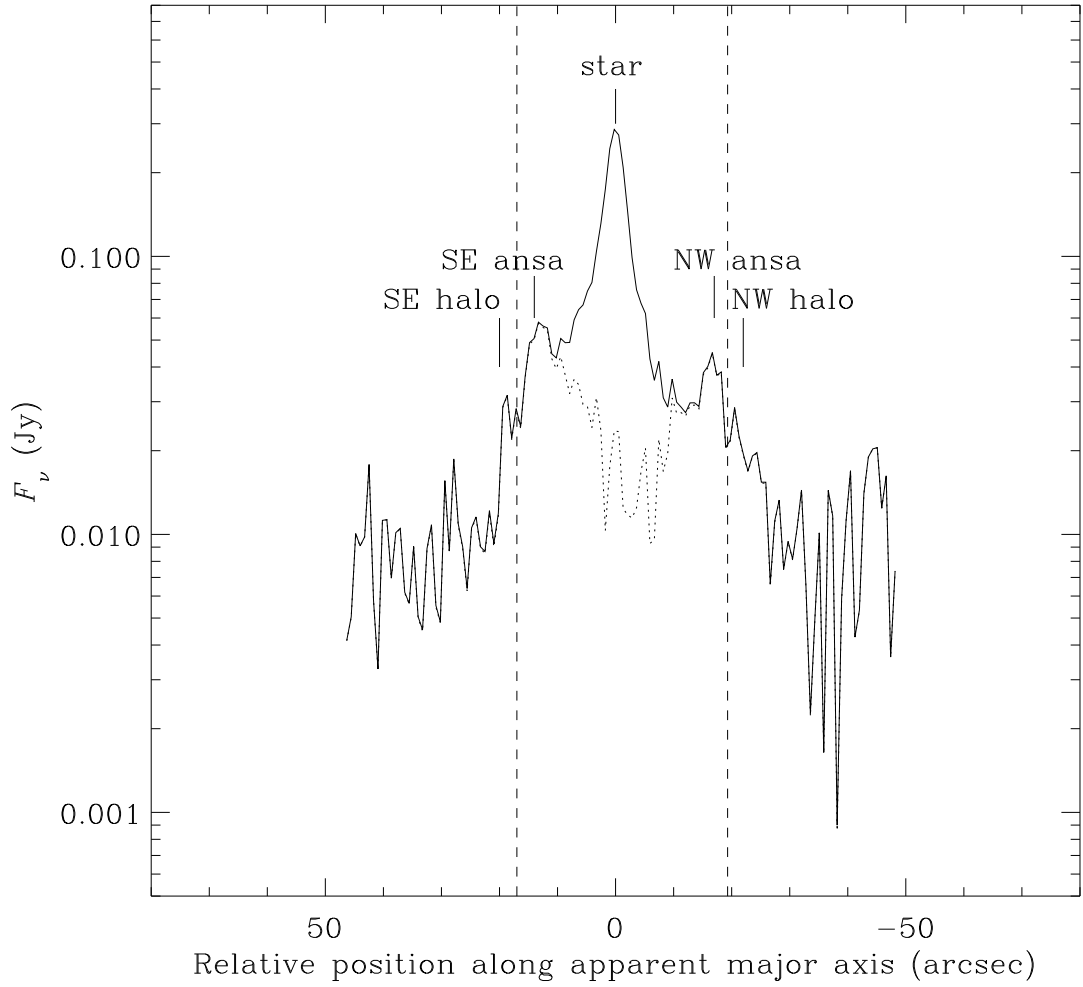}
\caption{Integrated brightness profile along the apparent major axis, where the positive direction corresponds to the SE direction, and the negative direction corresponds to the NW direction. The width for the integration was $23.^{\prime\prime}9$ (31 pixels). Profiles with (dotted line) and without (solid line) the PSF subtracted are shown. The edges of the ansai at $17^{\prime\prime}$ and $-19.3^{\prime\prime}$are noted with dashed lines.
\label{fig:profile}}
\end{figure} 

\newpage
\begin{figure}
\plotone{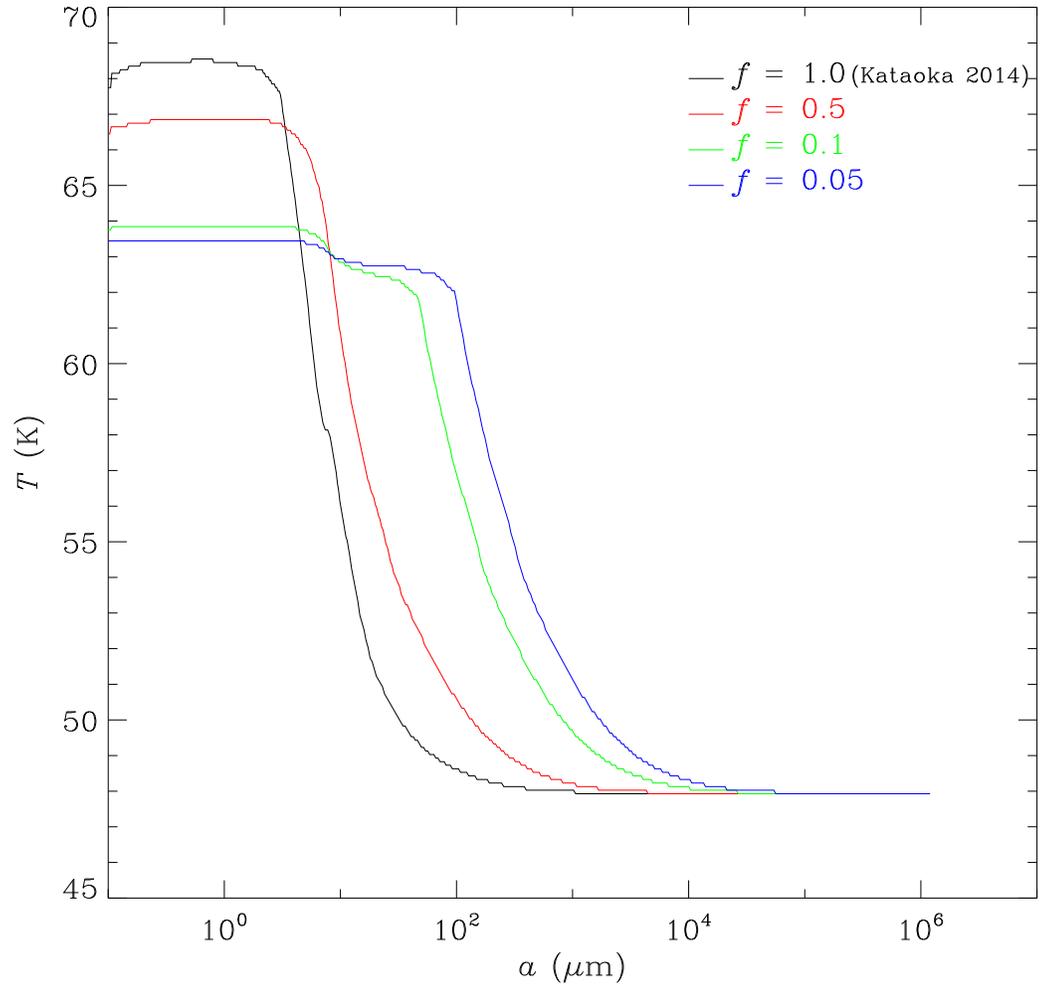}
\caption{Grain equilibrium temperatures $T(a)$ {\it vs}. grain radius $a$ for icy grains with volume filling factors $f=1.0$, 0.5, 0.1, and 0.05 at a distance of 140 AU.
\label{fig:temps}} 
\end{figure} 

\newpage
\begin{figure}
\plotone{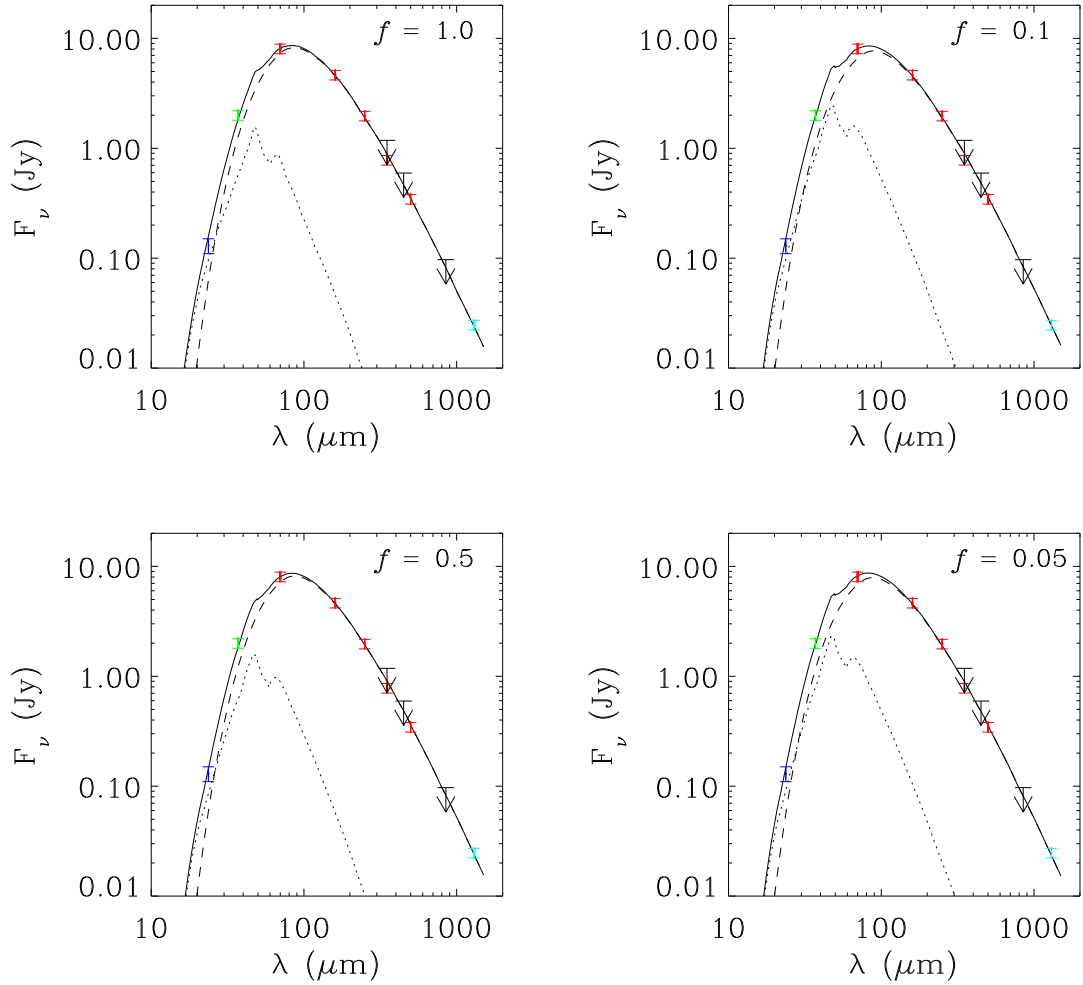}
\caption{Observed SED and best-fitting models for the icy aggregate grain compositions. The SEDs for icy grains over a range of porosities ($f = $ 1.0, 0.5, 0.1, and 0.05) \citep{kataoka14} are shown. The contributions from blowout grains (dotted lines) and large grains (dashed lines) are shown, with the total SED depicted as solid black lines. {\it Blue bar}: Spitzer/MIPS flux from photosphere-subtracted image \citep{stapelfeldt04}. {\it Green bar}: FORCAST data (this work). {\it Red bars}: Herschel model for the outer disk from \citet{acke12}. {\it Cyan bar:} ALMA data at 1.3 mm from \citet{macgregor17}.The submillimeter data are at 350 $\mu$m \citep{marsh05}, and 450 and 850 $\mu$m \citep{holland03}. The submillimeter fluxes are shown as upper limits because they include contamination from the inner disk. 
\label{fig:outer}}
\end{figure} 

\newpage
\begin{figure}
\plotone{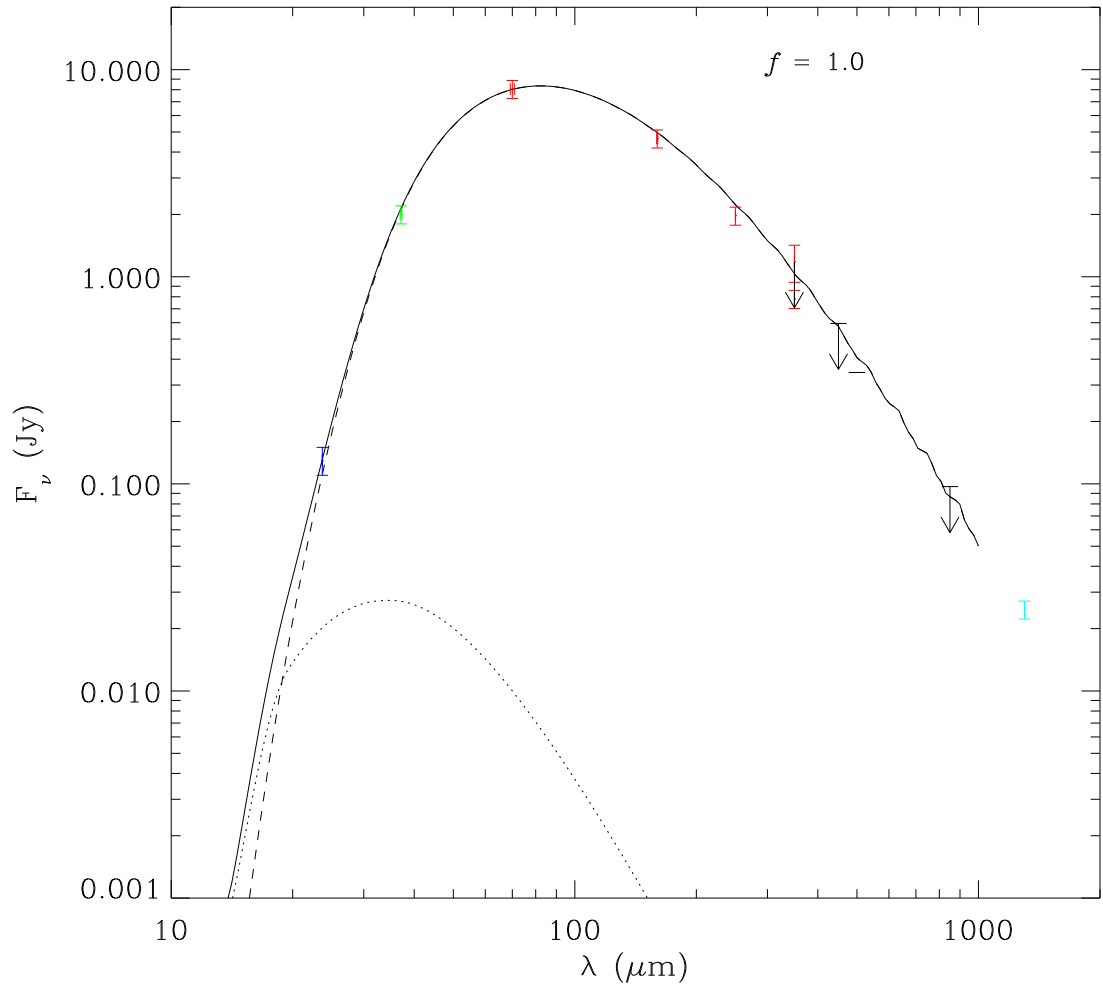}
\caption{Observed SED and model fits for astrosilicates. Lines and data points are the same as those in Figure \ref{fig:outer}.
\label{fig:outer_all}}
\end{figure} 

\newpage
\begin{figure}
\plotone{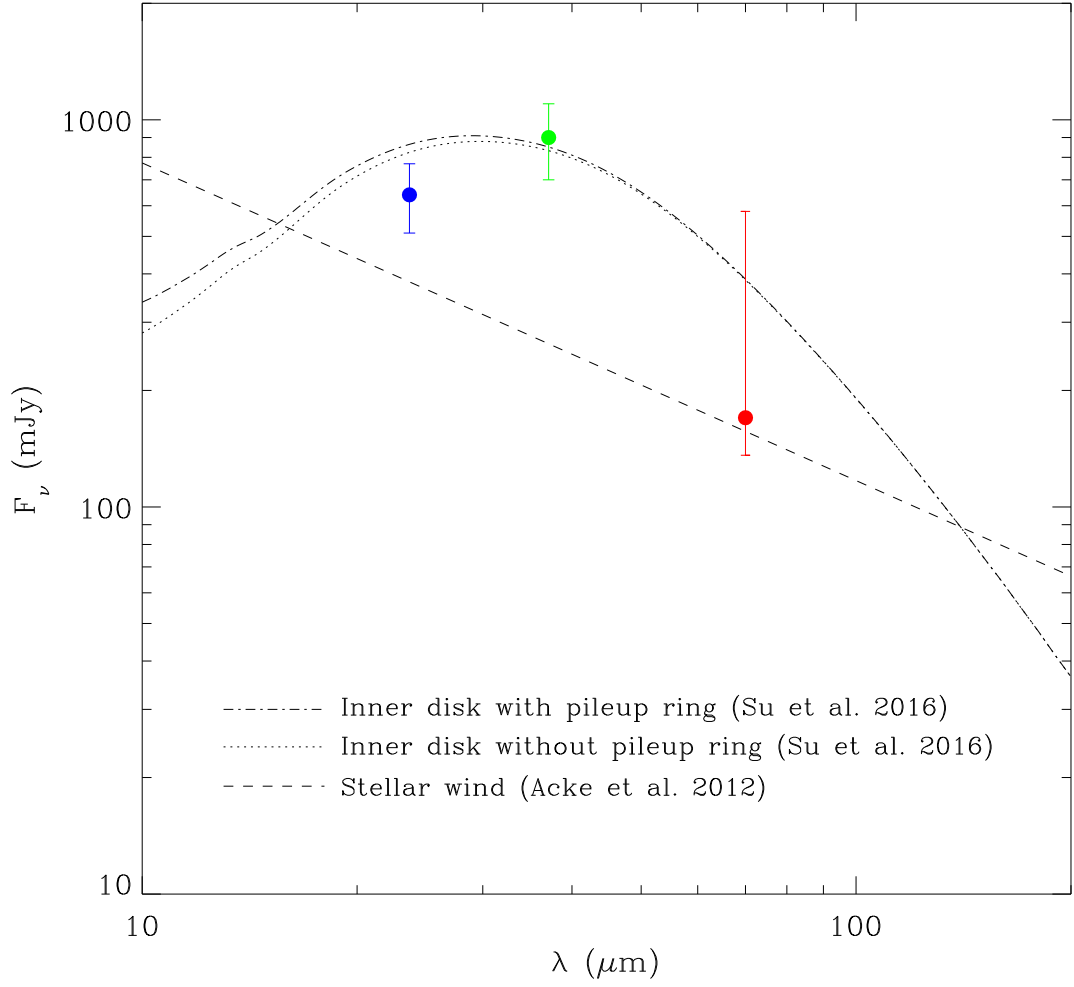}
\caption{Infrared excess {\it vs}. wavelength for the point source at 24 $\mu$m (blue point, Su et al. \citeyear{su16}), 37.1 $\mu$m (green point, this work), and 70 $\mu$m (red point, Acke et al. \citeyear{acke12}, Su et al. \citeyear{su13}). The total emission from the inner disk model of \citet{su16} is shown both with (dot-dashed line) and without (dotted line) the pileup ring. The expected infrared excess from a stellar wind (dahsed line) \citep{acke12} is shown.
\label{fig:ptsrc}}
\end{figure} 

\newpage
\floattable
\begin{deluxetable}{cccccc}
\tabletypesize{\tiny}
\tablecaption{\label{tab:obs} Summary of SOFIA/FORCAST observations of Fomalhaut under Plan ID $04\_0064$ and AOR ID $04\_0064\_1$.}
\tablehead{\colhead{Mission ID} & \colhead{Start time (UT)} & \colhead{End time (UT)} & \colhead{Altitude (ft)} & \colhead{Starting Elevation angle ($^\circ$)} & \colhead{Calibration uncertainty ($\%$)}}
\startdata
 2016-07-12$\_$FO$\_$F319 &  09:50:35 & 10:10:05 & 41000 & 30 & 6 \\
 2016-07-12$\_$FO$\_$F319 &  10:10:05 & 10:54:56 & 42000 & 33 & 6 \\
 2016-07-12$\_$FO$\_$F319 &  10:54:56 & 11:44:03 & 41000 & 39 & 6 \\
2016-07-14$\_$FO$\_$F321 &  10:49:47 & 11:22:55 & 38000 & 26 & 5 \\
2016-07-14$\_$FO$\_$F321 &  11:22:55 & 11:48:33 & 39000 & 33 & 5 \\
2016-07-14$\_$FO$\_$F321 &  11:48:33 & 12:16:41 & 39000 & 39 & 5 \\
2016-07-14$\_$FO$\_$F321 &  11:48:33 & 12:44:16 & 40000 & 46 & 5 \\
2016-07-14$\_$FO$\_$F321 &  12:44:27 & 13:31:10 & 41000 & 50 & 5 \\
2016-07-14$\_$FO$\_$F321 &  13:31:10 & 13:44:20 & 42000 & 57 & 6 \\
2016-07-14$\_$FO$\_$F321 &  13:58:29 & 14:10:08 & 42000 & 59 & 6 \\
2016-07-18$\_$FO$\_$F323 &  10:56:20 & 11:29:10 & 40000 & 25 & 5 \\
2016-07-18$\_$FO$\_$F323 &  11:29:10 & 11:59:41 & 40000 & 31 & 5 \\
2016-07-18$\_$FO$\_$F323 &  11:59:54 & 12:25:43 & 41000 & 38 & 5 \\
2016-07-18$\_$FO$\_$F323 &  12:25:43 & 12:51:40 & 42000 & 43 & 5 \\
2016-07-18$\_$FO$\_$F323 &  12:51:40 & 13:15:49 & 42000 & 49 & 5 \\
2016-07-20$\_$FO$\_$F325 &  10:07:12 & 10:46:56 & 39000 & 34 & 5 \\
2016-07-20$\_$FO$\_$F325 &  10:46:56 & 11:26:44 & 39000 & 41 & 5 \\
2016-07-20$\_$FO$\_$F325 &  11:26:55 & 12:04:21 & 39000 & 48 & 5 \\
2016-07-20$\_$FO$\_$F325 &  12:06:35 & 12:39:23 & 39000 & 52 & 5 \\
2016-07-20$\_$FO$\_$F325 &  12:41:21 & 13:16:46 & 41000 & 55 & 5 \\
\enddata
\end{deluxetable}

%\newpage
%\floattable
%\begin{deluxetable}{ccccccccccc}
%\tablecaption{\label{tab:outer2} Summary of silicate grain model \citep{pawellek14} parameters and results for the cold belt. Parameters are analogous to those in Table \ref{tab:outer}}
%\tablehead{\colhead{$\chi^2$} & \colhead{$q$} & \colhead{$a_{min}$ } & \colhead{$a_{max}$} & \colhead{$a_{blow}$} & \colhead{log $L_{d}$} & \colhead{log $M_{d}$} & \colhead{log $M_s$} & \colhead{log $\dot{M}$} &  \colhead{log $\dot{M}_{1\mu\rm{m}}$} \\
% & & ($\mu$m) & \colhead{(m)} & \colhead{($\mu$m)} & \colhead{(L$_\odot$)} & \colhead{(g)} & %\colhead{(g)} & \colhead{(g yr$^{-1}$)} & \colhead{(g yr$^{-1}$)}}
%\startdata
%\enddata
%\end{deluxetable}

\floattable
\newpage
\begin{center}
\rotate
\begin{deluxetable}{lcccccccccccccc}
\tabletypesize{\normalsize}
\tablecaption{\label{tab:outer} Summary of dust model parameters and results for the cold region. $M_d$ is the dust mass, $\dot{M}$ is the mass loss rate due to radiation pressure, $\dot{M}_{DP}$ is the scaled dust production rate for grains with $a_{min} \le a \le a_{blow}$ from \citet{kenyon08}, and $M_{1 \rm{mm}}$ is the mass in 1 -- 1000 $\mu$m (scaled with porosity) grains.}
\tablehead{ & & & \multicolumn{4}{c}{Large grains} & & \multicolumn{6}{c}{Blowout grains} & \\
\cline{4-7} \cline{9-14} \\
\colhead{Composition} & \colhead{$f$} & \colhead{$a_{blow}$} & \colhead{$q$} & \colhead{$a_{min}$ } & \colhead{$a_{max}$} & \colhead{log $M_{d}$} & & \colhead{$q$} & \colhead{$a_{min}$} & \colhead{$a_{max}$} & \colhead{log $M_{d}$} & \colhead{log $\dot{M}$} & \colhead{log $\dot{M}_{DP}$} & \colhead{$M_{1\rm{mm}}$} \\
 & & ($\mu$m) & & \colhead{($\mu$m)} & \colhead{(m)} & \colhead{(g)} &  & & \colhead{($\mu$m)} & \colhead{($\mu$m)} & \colhead{(g)} & \colhead{(g yr$^{-1}$)}  & \colhead{(g yr$^{-1}$)} & \colhead{(g)} } 
\startdata
icy aggregates & 1.0 & 10 & -3.4 & 12 & 0.04 & $26.78^{+0.14}_{-0.02}$ & & -3.5 & 0.1 & 12 & $23.62^{+0.20}_{-0.02}$ & $20.96^{+0.02}_{-0.11}$ & [20.65, 21.11] & $25.77^{+0.03}_{-0.03}$ \\
icy aggregates & 0.5 & 21 & -3.4 & 40 & 0.1 & $26.97^{+0.04}_{-0.03}$ & & -3.5 & 0.1 & 40 & $23.78^{+0.28}_{-0.20}$ & $21.00^{+0.03}_{-0.19}$ & [20.66, 21.12] & $25.83^{+0.04}_{-0.03}$\\
icy aggregates & 0.1 & 100 & -3.4 & 280 & 0.7 & $27.07^{+0.02}_{-0.03}$ & & -3.5 & 0.1 & 280 & $24.02^{+0.05}_{-0.11}$ & $21.17^{+0.04}_{-0.11}$ & [20.68, 21.14] & $25.82^{+0.03}_{-0.03}$ \\
icy aggregates & 0.05 & 210 & -3.4 & 540 & 1.2 & $27.03^{+0.02}_{-0.04}$ & & -3.5 & 0.1 & 540 & $23.99^{+0.06}_{-0.18}$ & $21.15^{+0.07}_{-0.18}$ & [20.69, 21.15] & $25.84^{+0.02}_{-0.04}$ \\
astrosilicates & 1.0 & 5 & -3.6 & 5.0 & 0.001 & $25.82^{+0.03}_{-0.04}$ & & -3.6 & 0.01 & 5.0 & $\le 21.86$ & $\le 19.38$ & [20.14, 20.60] & $25.81^{+0.03}_{-0.04}$\\
\enddata
\end{deluxetable}
\end{center}


\newpage

\begin{thebibliography}{ }

\bibitem[Absil et al.(2009)]{absil09}
Absil, O., Mennesson, B., Le Bouquin, J.-B. et al. 2009, \apj, 704, 150

\bibitem[Acke et al.(2012)]{acke12}
Acke, B., Min, M., Dominik, C. et al. 2012, \aap, 540, 125

\bibitem[Aumann(1985)]{aumann85}
Aumann, H. H. 1985, \pasp, 97, 885

\bibitem[Castelli \& Kurucz(2004)]{castelli04}
Castelli, F. \& Kurucz, R. L. 2004, arXiv:0405087

%\bibitem[Davis \& Ryan(1990)]{davis90}
%Davis, D. R. \& Ryan, E. V. 1990, \icarus, 83, 156

%\bibitem[De Buizer et al.(2012)]{debuizer12}
%De Buizer, J. M., Morris, M. R., Becklin, E. E. et al. 2012, \apj, 749L, 23

\bibitem[Dohnanyi(1969)]{dohnanyi69}
Dohnanyi, J. S. 1969, J. Geophys. Res., 74, 2531

\bibitem[Draine(2003)]{draine03}
Draine, B. T. 2003, \araa, 41, 241

\bibitem[Fraundorf et al.(1982)]{fraundorf82}
Fraundorf, P., Walker, R. M., \& Brownlee, D. E. 1982, in Comet Discoveries, Statistics, and Observational Selection, ed. L. L. Wilkening, IAU Colloq. 61, 383

\bibitem[Graham, Kalas \& Matthews(2007)]{graham07}
Graham, J. R., Kalas, P. G. \& Matthews, B. C. 2007, \apj, 654, 595

\bibitem[Griffin et al.(2010)]{griffin10}
Griffin, M. J., Abergel, A., Abreu, A. et al. 2010, \aap, 518L, 3

\bibitem[Harwit(1988)]{harwit88}
Harwit, M. 1998, Astrophysical Concepts (2nd ed.;  New York: Springer-Verlag)
\bibitem[Herter et al.(2012)]{herter12}
Herter, T. L., Adams, J. D., De Buizer, J. M. et al. 2012, \apjl, 749, L18

\bibitem[Herter et al.(2013)]{herter13}
Herter, T. L., Vacca, W. D., Adams, J. D. et al. 2013, \pasp, 125, 1393

\bibitem[Holland et al.(2003)]{holland03}
Holland, W. S., Greaves, J. S., Dent, W. R. F. et al. 2003, \apj, 582, 1141

\bibitem[Kalas, Graham, \& Clampin(2005)]{kalas05}
Kalas, P., Graham, J. R., \& Clampin, M. 2005, Nature, 435, 1067

\bibitem[Kalas et al.(2013)]{kalas13}
Kalas, P., Graham, J. R., Fitzgerald, M. P., \& Clampin, M. 2013, \apj, 775, 56

\bibitem[Kataoka et al.(2014)]{kataoka14}
Kataoka, A., Okuzumi, S., Tanaka, H. et al. 2014, \aap, 568, 42

\bibitem[Kenyon \& Bromley(2008)]{kenyon08}
Kenyon, S. J. \& Bromley, B. C. 2008, \apjs, 179, 451

\bibitem[Lau et al.(2013)]{lau13}
Lau, R. M., Herter, T. L., Morris, M. R. et al. 2013, \apj, 775, 37

\bibitem[Lebreton et al.(2013)]{lebreton13}
Lebreton, J., van Lieshout, R., Augereau, J.-C. et al. 2013, \aap, 555, A146

\bibitem[MacGregor et al.(2017)]{macgregor17}
MacGregor, M. A., Matr\`{a}, L., Kalas, P. et al. 2017, \apj, 842, 8

\bibitem[Mamajek(2012)]{mamajek12}
Mamajek, E. E. 2012, \apjl, 754, L20

\bibitem[Marsh et al.(2005)]{marsh05}
Marsh, K. A., Velusamy, T., Dowell, C. D. et al. 2005, \apjl, 620, L47

\bibitem[Min et al.(2010)]{min10}
Min, M., Kama, M., Dominik, C., \& Waters, L. B. F. M. 2010, \aap, 509, L6

%\bibitem[Min et al.(2011)]{min11}
%Min, M. Dullemond, C. P., Kama, M. et al. 2011, \icarus, 212, 416

\bibitem[Panagia \& Felli(1975)]{panagia75}
Panagia, N. \& Felli, M. 1975, \aap, 39, 1

\bibitem[Pawellek et al.(2014)]{pawellek14}
Pawellek, N., Krivov, A., Marshall, J. et al. 2014, \apj, 792, 65

\bibitem[Poglitsch et al.(2010)]{poglitsch10}
Poglitsch, A., Waelkens, C., Geis, N. et al. 2010, \aap, 518, L2

\bibitem[Pollack et al.(1994)]{pollack94}
Pollack, J. B., Hollenback, D., Beckwith, S. et al. 1994, \apj, 421, 615

\bibitem[Stapelfeldt et al.(2004)]{stapelfeldt04}
Stapelfeldt, K. R., Holmes, E. K., Chen, C. et al. 2004, \apjs, 154, 458

%\bibitem{Stern \& Colwell(1997)]{stern97}
%Stern, S. A. \& Colwell, J. E. 1997, \aj, 114, 841

\bibitem[Su et al.(2013)]{su13}
Su, K. Y. L., Rieke, G. H., Malhotra, R. et al. 2013, \apj, 763, 118

\bibitem[Su et al.(2016)]{su16}
Su, K. Y. L., Rieke, G. H., Defr\'{e}re, D. et al. 2016, \apj, 818, 45

\bibitem[Su et al.(2017)]{su17}
Su, K. Y. L., De Buizer, J. M., Rieke, G. H. et al. 2017, \aj, 153, 226

\bibitem[Tazaki \& Nomura(2015)]{tazaki15}
Tazaki, R. \& Nomura, N. 2015, \apj, 799, 119

%\bibitem[Thebault(2016)]{thebault16}
%Thebault, P. 2016, \aap, 587, 88

\bibitem[van Leeuwen(2007)]{vanleeuwen07}
van Leeuwen, F. 2007, \aap, 474, 653

%\bibitem[Wyatt, Clarke \& Booth(2011)]{wyatt11}
%Wyatt, M. C., Clarke, C. J. \& Booth, M. 2011, CeMDA, 111, 1 

\bibitem[Wolf \& Hillenbrand(2005)]{wolf05}
Wolf, S. \& Hillenbrand, L. A. 2005, Comp. Phys. Comm., 171, 208

\bibitem[Young et al.(2012)]{young12}
Young, E. T., Becklin, E. E., De Buizer, J. M. et al. 2012, \apjl, 749, L17

\end{thebibliography}
\end{document}